\documentclass[preprintnumbers,nofootinbib,preprint,12pt,showpacs]{revtex4}

\usepackage{amssymb,epsf}
\usepackage{graphicx}
\usepackage{latexsym}

\newcommand{\hM}{{\hat M}}
\newcommand{\hJ}{{\hat J}}
\newcommand{\vphi}{\varphi}
\newcommand{\sqdetg}{\sqrt{-g}}

\begin{document}

{~}

\vspace{1cm}

\title{Extremal Charged Rotating Black Holes in Odd Dimensions}

 \vspace{1.5truecm}
\author{
{\bf Masoud Allahverdizadeh, Jutta Kunz}
}
\affiliation{
{
Institut f\"ur Physik, Universit\"at Oldenburg, Postfach 2503\\
D-26111 Oldenburg, Germany}
}
\author{
{\bf Francisco Navarro-L\'erida}
}
\affiliation{
{Dept.~de F\'{\i}sica At\'omica, Molecular y Nuclear, Ciencias F\'{\i}sicas\\
Universidad Complutense de Madrid, E-28040 Madrid, Spain}
}

\begin{abstract}

Employing higher order perturbation theory,
we obtain charged rotating black holes in odd dimensions,
where the Einstein-Maxwell Lagrangian may be supplemented
with a Chern-Simons term. 
Starting from the Myers-Perry solutions,
we use the electric charge as the perturbative parameter,
and focus on extremal black holes with
equal-magnitude angular momenta.
For Einstein-Maxwell-Chern-Simons theory
with arbitrary Chern-Simons coupling constant,
we perform the perturbations up to third order for any odd dimension.
We discuss the physical properties of these black holes
and study their dependence on the charge.
In particular, we show that the gyromagnetic ratio $g$ 
of Einstein-Maxwell black holes differs from the
lowest order perturbative value $D-2$, and that the first correction term
to $g/(D-2)$ is universal.
\end{abstract}

\pacs{04.20.Ha, 04.20.Jb, 04.40.Nr, 04.50.-h, 04.70.Bw.}

\date{\today}

\maketitle 

\newpage

\section{Introduction}
String Theory, a major candidate for the quantum theory of gravity
and the unification of all interactions,
predicts that spacetime has more than four dimensions,
since it needs higher dimensions for its mathematical consistency.
String Theory has had considerable success in explaining the microscopic
degrees of freedom of black holes. In fact, the first successful
statistical counting of black hole entropy in string theory
was performed for a five-dimensional black hole \cite{Strominger:1996sh}.

Besides, higher-dimensional black hole spacetimes 
have become of high interest because of
various developments in gravity and high energy physics. 
In brane world models, i.e., scenarios
with large extra dimensions, for instance, 
one of the main predictions is the intriguing possibility
to produce higher-dimensional mini black holes in particle colliders. 
Such black holes may then provide a window into higher dimensions 
as well as into non-perturbative gravitational physics
which might already appear at the TeV scale. 
Naturally, one therefore would like to study and
understand the properties of such black holes 
corresponding to higher-dimensional generalizations
of the Kerr-Newman geometry.

While the non-rotating black hole solution in higher-dimensional
Einstein-Maxwell theory was found several decades ago 
\cite{Tangherlini:1963bw},
the counterpart of the Kerr-Newman solution in higher dimensions,
that is, the charged generalization of the Myers-Perry solution
\cite{Myers:1986un}
in higher dimensional Einstein-Maxwell theory, still
remains to be found analytically. 
However,
in the framework of supergravity theories and string theory
charged rotating black holes in higher dimensions 
have been obtained in closed form
\cite{Youm:1997hw,Emparan:2008eg,Kunz:2006jd}.
In particular, the general rotating black hole solution has
been obtained in the bosonic sector of minimal $D=5$ supergravity,
where the Einstein-Maxwell action is supplemented with a Chern-Simons term
\cite{Breckenridge:1996sn,Breckenridge:1996is,Cvetic:2004hs,Chong:2005hr}.

Lacking rotating Einstein-Maxwell black hole solutions in closed form,
basically two approaches have been followed to investigate their properties.
The first approach is a perturbative one.
Lowest order perturbation theory with
the charge as perturbative parameter \cite{Aliev:2004ec}
or
the angular momentum as perturbative parameter 
\cite{Aliev:2005npa,Aliev:2006yk}
has suggested, that the gyromagnetic ratio $g$
of Einstein-Maxwell black holes is simply given by $g=D-2$,
thus generalizing the Kerr-Newman value $g=2$.
The second approach is based on numerical and thus non-perturbative 
calculations of the black hole solutions
\cite{Kunz:2005nm,Kunz:2006eh}.
Here the $g$-factor was seen to deviate slightly from the perturbative value.

Recently, the perturbative approach has been extended to higher order.
With the charge as the perturbative parameter,
the properties of rotating black holes
in five dimensional Einstein-Maxwell theory have been
obtained
for the case of 
equal-magnitude angular momenta \cite{NavarroLerida:2007ez}.
There, in 3rd order perturbation theory,
the $g$-factor was shown to indeed deviate from 
the lowest order result $g=3$.

Here we generalize these results to 
extremal rotating black holes in
Einstein-Maxwell-Chern-Simons theory in 5 dimensions, 
allowing for an arbitrary value of the
Chern-Simons coupling constant $\tilde \lambda$.
For the supergravity value of $\tilde \lambda$
we recover the analytical solution up to this order,
thus showing the validity of the perturbative scheme,
while for vanishing $\tilde \lambda$ we recover the 
previous Einstein-Maxwell results \cite{NavarroLerida:2007ez}.

Then we generalize this perturbative approach
to Einstein-Maxwell-Chern-Simons black holes in higher odd dimensions.
Employing the charge as the perturbative parameter
we obtain the global and horizon properties of extremal
rotating black holes with equal-magnitude angular momenta
in 3rd order perturbation theory.
For the limiting case of pure Einstein-Maxwell theory
(where $\tilde \lambda=0$),
we show that the $g$-factor generically deviates from 
its lowest order perturbative value of $g=D-2$.
Interestingly, here the first correction term to $g/(D-2)$
is universal, i.e., independent of the dimension.

In section II we present the metric and the gauge potential
for black holes in odd dimensions with 
equal-magnitude angular momenta. 
We introduce the perturbation series 
and present the global and horizon properties of 
extremal black holes.
In section III we 
give the perturbative solution for
black holes of Einstein-Maxwell-Chern-Simons theory in 5 dimensions.
We then compare with the black hole solution of minimal $D=5$ supergravity,
known analytically,
and that of Einstein-Maxwell theory.
In section IV we present the 7-dimensional results
and the generalization to arbitrary odd dimensions.
We demonstrate the effect of the presence of charge on
the mass, the angular momentum, the magnetic moment
and the gyromagnetic ratio of these extremal rotating black holes.
We also evaluate their horizon properties.
We conclude in section V.
The formulae for the metric and the gauge potential
in $D$ dimensions are given in the Appendix.

\section{Black hole properties}

\subsection{Einstein-Maxwell and Einstein-Maxwell-Chern-Simons theory}

In odd-dimensional spacetimes, the 
Einstein-Maxwell Lagrangian
can be supplemented with a Chern-Simons term \cite{Gauntlett:1998fz},
leading to 
\begin{equation}
L = \frac{1}{16 \pi G_D} \sqdetg  \left(R - F_{\mu \nu} F^{\mu \nu} +
\frac{8}{D+1}{\tilde \lambda} \epsilon^{\mu_1 \mu_2 \dots
  \mu_{D-2}\mu_{D-1}\mu_D} F_{\mu_1 \mu_2} \dots F_{\mu_{D-2} \mu_{D-1}} A_{\mu_D}    \right) \ , \label{action}
\end{equation}
with curvature scalar $R$,
$D$-dimensional Newton constant $G_D$,
and field strength tensor
$ F_{\mu \nu} = \partial_\mu A_\nu -\partial_\nu A_\mu $,
where $A_\mu $ denotes the gauge potential. 
${\tilde \lambda}$ corresponds to the Chern-Simons coupling constant.
For vanishing ${\tilde \lambda}$ pure Einstein-Maxwell theory
is recovered.

The equations of motion are obtained by varying the action 
with respect to the metric $g_{\mu \nu}$, 
and the gauge potential $A_{\mu }$.
This leads to the Einstein equations
\begin{equation}
G_{\mu\nu}= R_{\mu\nu}-\frac{1}{2}g_{\mu\nu}R = 2 T_{\mu\nu}
\ , \label{FE1}
\end{equation}
with stress-energy tensor
\begin{equation}
T_{\mu \nu} = F_{\mu\rho} {F_\nu}^\rho - \frac{1}{4} g_{\mu \nu} F_{\rho
  \sigma} F^{\rho \sigma} \ ,
\end{equation}
and the gauge field equations
\begin{equation}
\nabla_\nu F^{\mu_1 \nu}  =  {\tilde \lambda}  \epsilon^{\mu_1 \mu_2 \mu_3 \dots
  \mu_{D-1}\mu_D} F_{\mu_2 \mu_3} \dots F_{\mu_{D-1} \mu_D}   \ .
\label{FE2}
\end{equation}

\subsection{Black holes in odd dimensions}

Stationary black holes in $D$ dimensions possess $N = [(D-1)/2]$
independent angular momenta $J_{i}$ associated with
$N$ orthogonal planes of rotation \cite{Myers:1986un}.
($[(D-1)/2]$ denotes the integer part of $(D-1)/2$,
corresponding to the rank of the rotation group $SO(D-1)$.)
The general black holes solutions then fall into two classes,
in even-$D$ and odd-$D$-solutions \cite{Myers:1986un}.
Here we focus on perturbative charged rotating black holes in odd dimensions.

When all $N$ angular momenta have equal-magnitude,
the symmetry of the solutions is strongly enhanced,
and the angular dependence factorizes 
in odd dimensions \cite{Kunz:2006eh}.
To obtain such perturbative charged generalizations of the
$D$-dimensional Myers-Perry solutions \cite{Myers:1986un}, 
we employ the following parametrization for the metric
\cite{Kunz:2006eh,Kunz:2006yp}

\begin{eqnarray}\label{metric1}
ds^2 =g_{tt}dt^2+\frac{dr^2}{W} + r^2\left[\sum^{N-1}_{i=1}\left(\prod^{i-1}_{j=0}\cos^{2}\theta_{j}\right)d\theta^{2}_{i}+\sum^{N}_{k=1}\left(\prod^{k-1}_{l=0}\cos^{2}\theta_{l}\right)\sin^{2}\theta_{k}d\varphi^{2}_{k}\right]
\nonumber \\
+V\left[\sum^{N}_{k=1}\left(\prod^{k-1}_{l=0}\cos^{2}\theta_{l}\right)\sin^{2}\theta_{k}\varepsilon_{k}d\varphi_{k}\right]^{2}-2B\sum^{N}_{k=1}\left(\prod^{k-1}_{l=0}\cos^{2}\theta_{l}\right)\sin^{2}\theta_{k}\varepsilon_{k}d\varphi_{k}dt \  ,
\end{eqnarray}
where $\theta_{0}\equiv0$, $\theta_{i}\in[0,\pi/2]$ 
for $i=1,...,N-1$, $\theta_{N}\equiv \pi/2$, 
$\varphi_{k}\in[0,2\pi]$ for $k=1,...,N$, 
and $\varepsilon_{k}=\pm1$ denotes the sense of rotation 
in the $k$-th orthogonal plane of rotation.
An adequate parametrization for the gauge potential is given by
\begin{eqnarray}\label{A1}
A_{\mu}dx^{\mu} &=& a_{0}+a_{\varphi}\sum^{N}_{k=1}\left(\prod^{k-1}_{l=0}\cos^{2}\theta_{l}\right)\sin^{2}\theta_{k}\varepsilon_{k}d\varphi_{k} \ , 
\end{eqnarray}
where the metric functions $g_{tt}$, $W$, $V$, $B$ 
and the functions for the gauge potential $a_{0}$,$a_{\varphi}$
depend only on the radial coordinate $r$.

\subsection{Perturbation theory} 

Before giving the perturbation series let us briefly
consider charge reversal symmetry.
In Einstein-Maxwell-Chern-Simons theory
in odd $D=2N+1$ dimensions, the Chern-Simons term contains
the gauge potential together with $N$ factors of the field strength tensor.
Thus a sign change of the gauge potential leads to a sign change
of the Chern-Simons term only when $N$ is even and not when $N$ is odd,
thus yielding the symmetry \cite{Kunz:2006yp}
\begin{eqnarray}
{\rm even\ } N (D=5,9,13,\dots):
&& \hspace{1cm} (Q,\tilde \lambda) \longrightarrow (-Q, -\tilde \lambda)
\ , \label{cs_sym1}
\\
{\rm odd\ } N (D=7,11,15,\dots):
&& \hspace{1cm} (Q,\tilde \lambda) \longrightarrow (-Q, +\tilde \lambda)
\ . \label{cs_sym2}
\end{eqnarray}
Thus when $N$ is odd, Einstein-Maxwell-Chern-Simons theory
possesses charge reversal symmetry just like Einstein-Maxwell theory,
whereas when $N$ is even, only the discrete symmetry Eq.~(\ref{cs_sym1}) holds.

We now consider perturbations about the Myers-Perry solutions,
with the electric charge as the perturbative parameter.
In the presence of a Chern-Simons term
we obtain the perturbation series in the general form
\begin{eqnarray}\label{gtt}
g_{tt} = -1+\frac{2\hat{M}}{r^{D-3}}+q^{2}g^{(2)}_{tt}
+q^{3}g^{(3)}_{tt} +O(q^{4}) \ ,
\end{eqnarray}
\begin{eqnarray}\label{W}
W =1-\frac{2\hat{M}}{r^{D-3}}+\frac{2\hat{J}^{2}}{\hat{M}r^{D-1}}
+q^{2}W^{(2)}+q^{3}W^{(3)}+O(q^{4}) \ ,
\end{eqnarray}
\begin{eqnarray}\label{N}
V = \frac{2\hat{J}^{2}}{\hat{M}r^{D-3}}+q^{2}V^{(2)}
+q^{3}V^{(3)}+O(q^{4}) \ ,
\end{eqnarray}
\begin{eqnarray}\label{B}
B = \frac{2\hat{J}}{r^{D-3}}+q^{2}B^{(2)}
+q^{3}B^{(3)}+O(q^{4}) \ ,
\end{eqnarray}
\begin{eqnarray}\label{a0}
a_{0} = q a^{(1)}_{0} +q^{2} a^{(2)}_{0}
+q^{3} a^{(3)}_{0}+O(q^{4}) \ ,
\end{eqnarray}
\begin{eqnarray}\label{avarphi}
a_{\varphi} = q a^{(1)}_{\varphi} +q^{2} a^{(2)}_{\varphi}
+q^{3} a^{(3)}_{\varphi}+O(q^{4}) \ ,
\end{eqnarray}
where $q$ is the perturbative parameter associated with the
electric charge (see Eq.~(\ref{charge}) below).
Since the Chern-Simons term does not modify the expression
for the energy-momentum tensor,
the expansion of the metric contains only quadratic and higher
order terms in the perturbative parameter.

Taking into account the symmetry with respect to charge
reversal for odd $N$ Einstein-Maxwell-Chern-Simons theory
as well as pure Einstein-Maxwell theory, 
the perturbations reduce in these cases to only even terms for the metric
and only odd terms for the gauge potential.
The presence of this symmetry
therefore implies a considerable simplification for higher
order perturbation theory. 

In even $N$ Einstein-Maxwell-Chern-Simons theory 
the symmetry Eq.~(\ref{cs_sym1}) 
has also implications for the perturbative expansion.
It follows, for instance, that 
in the perturbative expansion of the gauge potential
even powers in the perturbative parameter $q$ 
must be multiplied by odd powers of the
Chern-Simons coupling $\tilde \lambda$. 

Inserting the metric Eq.~(\ref{metric1}) and
the gauge potential Eq.~(\ref{A1})
with the perturbation expansion Eqs.~(\ref{gtt}-\ref{avarphi})
into the field equations Eqs.~(\ref{FE1}-\ref{FE2}),
we then solve these equations order by order.

\subsection{Physical Quantities }

The mass $M$, the angular momenta $J$, 
the electric charge $Q$, and the magnetic moments $\mu_{mag}$ 
can be read off the asymptotic behavior of the metric, and the
gauge potential \cite{Kunz:2006eh,Kunz:2006yp}
\begin{eqnarray}\label{quantities}
g_{tt}=-1+\frac{\tilde{M}}{r^{D-3}}+... \ ,  \quad B=\frac{\tilde{J}}{r^{D-3}}+... \ , \quad  a_{0}=\frac{\tilde{Q}}{r^{D-3}}+... \ ,  \quad   a_{\varphi}=\frac{\tilde{\mu}_{mag}}{r^{D-3}}+... \ ,
\end{eqnarray}
where
\begin{eqnarray}\label{quantities1}
\tilde{M}=\frac{16\pi G_{D}}{(D-2)A(S^{D-2})}M \ ,  
\nonumber \\
\tilde{J}=\frac{8\pi G_{D}}{A(S^{D-2})}J \ ,\nonumber \\ 
\tilde{Q}=\frac{4\pi G_{D}}{(D-3)A(S^{D-2})}Q \ ,   
\nonumber \\
\tilde{\mu}_{mag}=\frac{4\pi G_{D}}{(D-3)A(S^{D-2})}\mu_{mag} \ ,
\end{eqnarray}
and $A(S^{D-2})$ is the area of the unit $(D-2)$-sphere. 
The gyromagnetic ratio $g$ is given by
\begin{eqnarray}\label{g}
g = \frac{2M\mu_{mag}}{QJ}  \ .
\end{eqnarray}

The horizon angular velocities $\Omega$ 
can be defined by imposing the Killing vector field
\begin{eqnarray}\label{chi}
\chi = \xi+\Omega\sum^{N}_{k=1}\epsilon_{k}\eta_{k} \ ,
\end{eqnarray}
to be null on and orthogonal to the horizon. 
The horizon electrostatic potential $\Phi_{H}$ 
of these black holes is given by
\begin{eqnarray}\label{PhiH}
\Phi_{H} = \left. (a_{0}+\Omega a_{\varphi}) \right|_{r=r_{H}} \ ,
\end{eqnarray}
and the surface gravity $\kappa_{sg}$ is defined by
\begin{eqnarray}\label{sg1}
\kappa^{2}_{sg} = \left.
-\frac{1}{2}(\nabla_{\mu}\chi_{\nu})(\nabla^{\mu}\chi^{\nu})\right|_{r=r_{H}} \ .
\end{eqnarray}

The black holes further satisfy the Smarr-like mass formula 
\cite{Gauntlett:1998fz},
\begin{equation}
M = \frac{D-2}{D-3} \frac{\kappa_{sg} A_{H}}{8 \pi G_D} + \frac{D-2}{D-3} N \Omega
J  +  \Phi_{H} Q + \frac{D-5}{D-3} {\tilde \lambda} I \ , \label{smarr}
\end{equation}
where $I$ denotes the integral
\begin{equation}
I = -\frac{1}{4 \pi G_D} \int_\Sigma dS_\sigma \chi^\nu F_{\nu \rho} J^{\rho \sigma} \
, \label{int_I}
\end{equation}
and
$J^{\rho \sigma}$ is defined by
\begin{equation}
J^{\rho \sigma} = - \epsilon^{\rho \sigma \mu_1 \mu_2 \mu_3 \dots \mu_{D-3}
  \mu_{D-2}} A_{\mu_1} F_{\mu_2 \mu_3} \dots F_{\mu_{D-3} \mu_{D-2}} \ . \label{current} 
\end{equation}
Clearly, the Smarr-like relation Eq.~(\ref{smarr}) 
is special in 5 dimensions,
since the last term vanishes there,
independent of the value of the Chern-Simons coupling constant.
Thus in this case the mass relation Eq.~(\ref{smarr}) 
reduces to the Smarr relation of Einstein-Maxwell theory.

\section{Charged black holes in 5 dimensions}

Black holes in minimal $D=5$ supergravity are known exactly
\cite{Breckenridge:1996sn,Breckenridge:1996is,Cvetic:2004hs,Chong:2005hr}.
Here the Chern-Simons coupling constant $\tilde \lambda$ assumes the value
$\tilde \lambda_{SG} = \frac{1}{2 \sqrt{3}}$.
For convenience, we therefore here redefine the 
Chern-Simons coupling constant as
\begin{equation}
\lambda = 2 \sqrt{3} \tilde \lambda \ .
\end{equation}
Thus $\lambda_{SG}=1$. Moreover, we will set $G_D=1$ in the following.

The rotating black holes of $D=5$ supergravity have intriguing properties
\cite{Gauntlett:1998fz,Kunz:2005ei}.
For general $\lambda > \lambda_{SG}$, these black hole solutions
were studied numerically \cite{Kunz:2005ei}.
Here further surprising properties 
manifest themselves, such as, for instance, non-uniqueness.
Interestingly, the numerical studies also showed,
that supersymmetry
represents the borderline between stability and instability
\cite{Kunz:2005ei}.

Recently, 
perturbative solutions of Einstein-Maxwell-Chern-Simons
black holes were obtained for arbitrary
Chern-Simons coupling constant $\lambda$,
by calculating the solutions to lowest order in 
the two independent angular momenta \cite{Aliev:2008bh}.

Here we first present the perturbative solutions,
obtained in 3rd order in the charge,
for extremal black holes with equal-magnitude angular momenta
in Einstein-Maxwell-Chern-Simons theory
for arbitrary Chern-Simons coupling constant $\tilde \lambda$.
Then we consider the special case of 
minimal $D=5$ supergravity,
in order to compare with the exact solution.
Last we recover the results of \cite{NavarroLerida:2007ez}
obtained in pure Einstein-Maxwell theory.

\subsection{Einstein-Maxwell-Chern-Simons black holes}

Beside the discrete symmetry
with respect to charge reversal Eq.~(\ref{cs_sym1})
we have also a discrete symmetry with respect to the sense of rotation
of the equal-magnitude angular momenta $J_k=\varepsilon_k J$.
In 5 dimensions we have two angular momenta $J_1$ and $J_2$
and the associated discrete symmetry is
\begin{equation}
(\varepsilon_1 \varepsilon_2, \lambda) 
\longrightarrow (-\varepsilon_1 \varepsilon_2, -\lambda) .
\end{equation}
Consequently, a solution with $J_1=-J_2$
is equivalent to the solution with $J_1=J_2$ 
and with the same electric charge
but with the opposite value of $\lambda$.
It is therefore sufficient, to consider
perturbation theory only for the case $J_1=J_2=J$.

We now perform perturbation theory in the charge to 3rd order,
while we retain the Chern-Simons coupling constant $\lambda$
as an arbitrary constant.
We impose the extremality condition on the black holes 
and fix the angular momentum $J$ for all orders.
We further impose a regularity condition at horizon. 
This choice then fixes all integration constants
which appear in the scheme.

Introducing the parameter $\nu$ for the extremal
Myers-Perry solutions by
\begin{equation}
\hM=2\nu^2, \quad\quad
\hJ=2\nu^3, 
\label{nu5}
\end{equation}
we obtain for the metric 
the perturbation expansion up to 3rd order
in the perturbative parameter $q$
and for the gauge functions up to 4th order 
\begin{eqnarray}
g_{tt} &=& -1 + \frac{4\nu^2}{r^2} +\frac{r^2-4\nu^2}{3\nu^2 r^4} q^2
+ \frac{\lambda \sqrt{3}}{27\nu^4 r^2}q^3 +O(q^4) \ ,
    \nonumber \\
W&=&1-\frac{4\nu^2}{r^2}+\frac{4 \nu^4}{r^4} -\frac{r^2-2\nu^2}{3\nu^2
    r^4}q^2 - \frac{\lambda \sqrt{3} (r^2-2\nu^2)}{27\nu^4 r^4}q^3 + O(q^4) \ ,
    \nonumber \\
V&=&\frac{4\nu^4}{r^2} -\frac{2(r^2+2\nu^2)}{3 r^4} q^2 
   + \frac{2 \lambda \sqrt{3} (r^2+6\nu^2)}{27\nu^2 r^4} q^3 + O(q^4) \ ,
    \nonumber \\
B&=&\frac{4 \nu^3}{r^2} - \frac{4\nu}{3 r^4}  q^2 
+ \frac{2\lambda \sqrt{3}}{9\nu r^4} q^3 
+O(q^4) \ ,
\nonumber \\
a_t &=& \frac{1}{r^2}q + \frac{(1-\lambda^2)}{9\nu^6 r^4} 
\left[2\nu^2 (r^2-\nu^2) +
  r^2(r^2-2\nu^2)\ln\left(1-\frac{2\nu^2}{r^2}\right)\right] \left(q^3 -
  \frac{\lambda \sqrt{3}}{3\nu^2} q^4 \right) + O(q^5) \ , \ \ \ \ 
\nonumber \\
a_\vphi&=&-\frac{\nu}{r^2}q + \frac{\lambda\sqrt{3}}{6\nu r^2} q^2 -
  \frac{(1-\lambda^2)}{18 \nu^7 r^4}\left[\nu^2(2 r^4+ \nu^2 r^2 - 4 \nu^4) +
    r^2 (r^4 -4 \nu^4) \ln\left(1-\frac{2\nu^2}{r^2}\right)\right] q^3
  \nonumber \\
&&+\frac{\lambda\sqrt{3}}{54 \nu^7 r^4} \left[-\frac{\lambda^2\nu^2}{4}r^2 +
  2\nu^2(1-\lambda^2)(4 r^2-3\nu^2) +
  4(1-\lambda^2)r^2(r^2-2\nu^2)\ln\left(1-\frac{2\nu^2}{r^2}\right)\right]q^4
\nonumber \\
&&+ O(q^5) \ . \label{solution_EMCS}
\end{eqnarray}
We note, that the gauge function $a_t$ contains no quadratic term
in the perturbative parameter.

From these expansions the physical properties
of the black hole solutions can be extracted.
To third order the global physical quantities are given by
\begin{eqnarray}
&&M=\frac{3}{2}\pi \nu^2 + \frac{\pi}{8\nu^2} q^2 
+\frac{\lambda\sqrt{3}}{72\nu^4}\pi q^3 
 + O(q^4) \ , \nonumber \\
&& J=\pi \nu^3 \ (\mbox{for any order}) \ , \ \ \ Q=\pi q \ , 
\nonumber \\
&&\mu_{\rm {mag}}=\pi \nu q -\frac{\lambda\sqrt{3}}{6\nu}\pi q^2 -
\frac{(1-\lambda^2)}{18 \nu^3} \pi q^3 + \frac{\lambda^3\sqrt{3}}{216 \nu^5}\pi q^4 + O( q^5) \
, \label{mag_mom_EMCS}
\end{eqnarray}
yielding the gyromagnetic ratio $g$ 
for general Chern-Simons coupling constant $\lambda$
\begin{equation}
g=
3 -\frac{\lambda\sqrt{3}}{2\nu^2}q + \frac{(1+2\lambda^2)}{12 \nu^4} q^2 
 -\frac{(1-\lambda^2)\lambda\sqrt{3}}{72 \nu^6}q^3 + O(q^4) \ . \label{g_factor_extremal_EMCS}
\end{equation}

\begin{figure}[tbp]
\epsfxsize=7cm
\centerline{\includegraphics[angle=0,width=92mm]{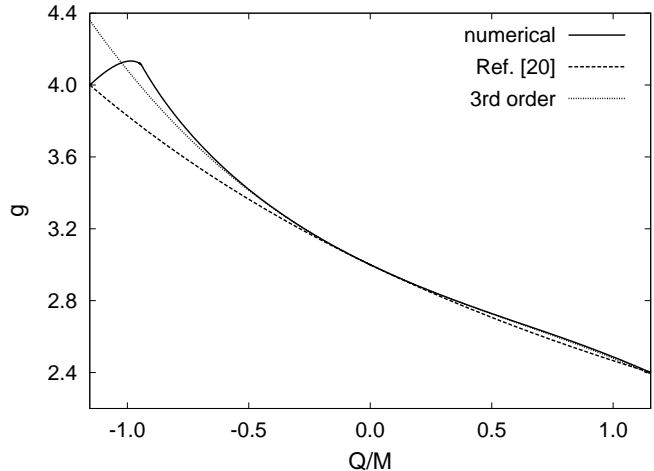}}
\caption{
The gyromagnetic ratio $g$ versus the 
charge to mass ratio $Q/M$
for $D = 5$ Einstein-Maxwell-Chern-Simons black holes
for Chern-Simons coupling constant $\lambda=1/2$.
Exact numerical values are shown by the solid line,
1st order perturbative values \cite{Aliev:2008bh} by the dashed line,
and the present 3rd order perturbative values by the dotted line.
}
\label{fig1}
\end{figure}

As an example, we exhibit the 3rd order perturbative result
for the $g$-factor versus the charge to mass ratio $Q/M$
for the extremal Einstein-Maxwell-Chern-Simons black holes
with Chern-Simons coupling constant $\lambda=1/2$
in Fig.~\ref{fig1}.
For comparison, we also exhibit the corresponding
numerical results \cite{Kunz:2005ei}
and the lowest order results
of Aliev and Ciftci \cite{Aliev:2008bh},
where perturbation theory is performed in the angular momenta.

The allowed range of the charge to mass ratio $Q/M$
is bounded by the extremal Tangherlini values,
$-2/\sqrt{3} \le Q/M \le 2/\sqrt{3}$,
as long as $\lambda \le 1$ \cite{Kunz:2005ei}.
The figure shows, that
the 3rd order perturbative results for the $g$-factor 
are in excellent agreement with the numerical results
for a large range of values.
Only toward the lower end of the interval significant deviations arise.
The first order results of \cite{Aliev:2008bh}
exhibit larger deviations over most of the interval,
with the exception of the central region
and the endpoints, where the agreement with the numerical results is perfect.

The event horizon is located at $r=r_{H}$, where
\begin{equation}
r_{H} = \sqrt{2} \nu + \frac{\sqrt{2}}{24 \nu^3} q^2 
+\frac{\lambda \sqrt{2}}{36 \nu^5} q^3
+ O( q^4) \ . \label{hor_rad_EMCS}
\end{equation}
The horizon properties are given by
\begin{eqnarray}
&&\Omega = \frac{1}{2\nu} - \frac{1}{24 \nu^5} q^2 
-\frac{\lambda\sqrt{3}}{108 \nu^7}q^3
+ O(q^4) \ , 
\nonumber \\
%
&&A_{H} = 8 \pi^2 \nu^3 +\frac{2\pi^2\lambda\sqrt{3}}{9\nu^3}q^3 + O(q^5) \ ,
\nonumber \\
&&\Phi_{H}=\frac{1}{4\nu^2} q 
+\frac{\lambda\sqrt{3}}{24\nu^4}q^2
+ \frac{(1-\lambda^2)}{72 \nu^6} q^3 
-\frac{5\lambda^3\sqrt{3}}{864\nu^8}q^4
+O(q^5) \ , \label{AH_and_PhiH_EMCS}
\end{eqnarray}
and the surface gravity $\kappa_{\rm sg}$
vanishes for these extremal solutions.

The expressions for $a_t$ and $a_\vphi$ in Eq.~(\ref{solution_EMCS}) 
(and also for the higher order terms of the metric functions not shown here)
reveal that the expansions contain, in general, logarithmic terms. 
However, these logarithmic terms disappear for $|\lambda|=1$,
and thus for the supergravity case. 

\subsection{Special cases}

To address the validity of the perturbative results,
we now specialize to the $D=5$ supergravity value of
the Chern-Simons coupling constant, $\lambda=1$,
since here the exact solutions are available for comparison.
Afterwards we consider pure Einstein-Maxwell theory,
where $\lambda=0$.

\subsubsection{Supergravity: $\lambda=1$}

For $\lambda=1$ the
exact solutions are known analytically 
for general values of the angular momenta \cite{Chong:2005hr}.
For equal angular momenta the non-extremal solutions are given by
\begin{eqnarray}
&&g_{tt} = -1 + \frac{2 m }{r^2} - \frac{4 q^2}{3 r^4} \ ,
\nonumber \\
&&W = 1 - \frac{2 m}{r^2} + \frac{2(2 q^2 + 2 \sqrt{3} q a^2 + 3 a^2 m)}{3
  r^4}  \ ,
\nonumber \\
&&V = \frac{2 a^2(2\sqrt{3} q + 3 m)}{3 r^2}  - \frac{4 q^2 a^2}{3 r^4}\ ,
\nonumber \\
&&B = \frac{2 a(\sqrt{3} q + 3 m)}{3 r^2} - \frac{4 q^2 a}{3 r^4}  \ ,
\nonumber \\
&&a_t = \frac{q}{r^2}\ ,
\nonumber \\
&&a_\vphi = - \frac{q a}{r^2} \ . \label{exact_supergravity}
\end{eqnarray}

The extremal solutions which are continuously connected 
to the Myers-Perry solutions
in the limit of vanishing electric charge ($q=0$) are characterized by
\begin{equation}
m = \frac{2(\sqrt{3}q + 3 a^2)}{3} \ , \label{extremality_supergravity}
\end{equation}
with $\sqrt{3}q + 3 a^2 \ge 0$.

It is easy to see that Eq.~(\ref{exact_supergravity}), when restricted to the
extremal case by Eq.~(\ref{extremality_supergravity}), 
give rise to expansions which 
precisely coincide with the perturbative expansions
Eq.~(\ref{AH_and_PhiH_EMCS}) for $\lambda=1$. 
One has just to redefine the
rotation parameter $a$ by 
\begin{equation}
a(\sqrt{3} q + 2 a^2)=2 \nu^3 \ , \label{redef_J}
\end{equation}
and use the series expansion of $a$ 
in terms of $q$ and $\nu$ to the appropriate order
\begin{equation}
a = \nu - \frac{\sqrt{3}}{6\nu} q + \frac{\sqrt{3}}{216 \nu^5}q^3 +
\frac{1}{432\nu^7}q^4 + O(q^6) \ .  \label{expansion_a}
\end{equation}

The global properties of the exact non-extremal supergravity
solutions are given by
\begin{eqnarray}
&&M=\frac{3\pi m}{4} \ , \nonumber \\
&&J=\frac{\pi a (\sqrt{3}q + 3 m)}{6} \ , \ \ \ Q=\pi q \ , \nonumber \\
&&\mu_{\rm {mag}}= \pi q a \ , \nonumber \\
&&g=\frac{9 m}{\sqrt{3} q + 3 m} \ . \label{charges_EMCS_supergravity}
\end{eqnarray}
The corresponding extremal expressions are then obtained by substituting
Eq.~(\ref{extremality_supergravity}) and Eq.~(\ref{redef_J}) in
Eq.~(\ref{charges_EMCS_supergravity}). 

In particular, the gyromagnetic ratio of these supergravity black holes
can be expressed in terms of the charge to mass ratio $Q/M$
\begin{equation}
g=\frac{3}{\frac{\sqrt{3}}{4} \frac{Q}{M} + 1} ,
\end{equation}
for non-extremal and extremal black holes alike.
Thus the $g$-factor 
depends only on the charge to mass ratio $Q/M$.
For the whole set of equal-magnitude 
angular momenta black holes (whether extremal or non-extremal)
the $g$-factor therefore 
reduces to a single line.

\begin{figure}[tbp]
\epsfxsize=7cm
\centerline{\includegraphics[angle=0,width=92mm]{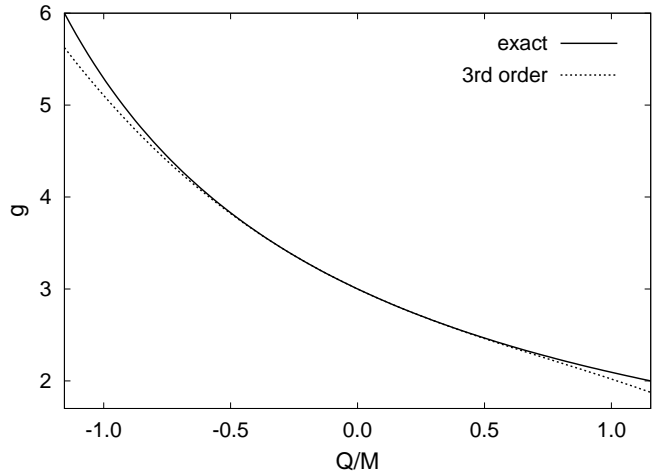}}
\caption{
The gyromagnetic ratio $g$ versus the 
charge to mass ratio $Q/M$
for $D = 5$ Einstein-Maxwell-Chern-Simons black holes
for Chern-Simons coupling constant $\lambda=1$. 
Exact values are shown by the solid line 
and 3rd order perturbative values by the dotted line. 
}
\label{fig2}
\end{figure}

Clearly, this $g$-factor is a monotonic function of 
the charge to mass ratio $Q/M$. 
When $Q/M$ varies in the allowed range
$-2/\sqrt{3} \le Q/M \le 2/\sqrt{3}$,
bounded by the extremal Tangherlini values,
the gyromagnetic ratio varies in the range
$6 \ge g \ge 2$.
Note, that for opposite angular momenta,
the gyromagnetic ratio varies in the opposite way,
i.e., $2 \le g \le 6$.
The exact gyromagnetic ratio for black holes with equal angular momenta
is exhibited in Fig.~\ref{fig2}.

To obtain an estimate of the range of validity
of the perturbative results presented above, we 
express the 3rd order perturbative $g$-factor 
Eq.~(\ref{g_factor_extremal_EMCS})
in terms of the charge to mass ratio $Q/M$ for $\lambda=1$
\begin{equation}
g= 3 -\frac{3 \sqrt{3}}{4} \frac{Q}{M}
+ \frac{9}{16} \left( \frac{Q}{M} \right)^2
- \frac{9 \sqrt{3}}{64} \left( \frac{Q}{M} \right)^3
+ \dots \ ,
\label{g_3cs}
\end{equation}
and exhibit this perturbative result for the $g$-factor
also in Fig.~\ref{fig2}.
Comparison shows that the 3rd order perturbation theory results 
are excellent for a large range of values.
Only towards the ends of the interval deviations arise. 

It is amusing to note, that lowest order perturbation theory
in the angular momenta, as performed by Aliev and Ciftci
\cite{Aliev:2008bh},
even yields the exact result for the whole range of $Q/M$.
The reason is that, as seen above, for equal-magnitude angular momenta
the value of $g$ depends only on the charge to mass ratio
and not on the angular momentum.
In this sense, the gyromagnetic ratio
for the case of $D=5$ supergravity is very special
for black holes with equal-magnitude angular momenta.
For general Chern-Simons coupling, however, this is no
longer the case. Here the values of $g$ do depend on the
charge to mass ratio as well as on the
angular momentum.

\subsubsection{Pure Einstein-Maxwell black holes}

When we set the Chern-Simons coupling constant to zero,
we obtain the perturbative results for Einstein-Maxwell black holes
\begin{eqnarray}
&&g_{tt}=-1+\frac{4\nu^2}{r^2} 
+\frac{r^2-4\nu^2}{3\nu^2 r^4} q^2 
+O(q^4) \ ,
    \nonumber \\
&&W=1-\frac{4\nu^2}{r^2}+\frac{4 \nu^4}{r^4} -\frac{r^2-2\nu^2}{3\nu^2
    r^4}q^2
+O(q^4) \ ,
    \nonumber \\
&&V=\frac{4\nu^4}{r^2} -\frac{2(r^2+2\nu^2)}{3 r^4} q^2  
+O(q^4) \ ,
\nonumber \\
&&B=\frac{4 \nu^3}{r^2} - \frac{4\nu}{3 r^4}  q^2 
+O(q^4) \ ,
\nonumber \\
&& a_t=\frac{1}{r^2} q  + \left[\frac{2(r^2-\nu^2)}{9\nu^4 r^4}
   + \frac{r^2-2\nu^2}{9 \nu^6 r^2} \ln\left(1-\frac{2
    \nu^2}{r^2}\right)\right] q^3  + O(q^5) \ , \nonumber \\
&&a_\vphi=-\frac{\nu}{r^2}  q - \left[ \frac{2 r^4 + \nu^2 r^2 -4 \nu^4}{18
    \nu^5 r^4} + \frac{r^4-4 \nu^4}{18 \nu^7 r^2} \ln\left(1-\frac{2
    \nu^2}{r^2}\right) \right] q^3 + O(q^5) \ ,  \label{solution}
\end{eqnarray}
which coincides to this order
with the results of \cite{NavarroLerida:2007ez}.

The global physical quantities are given by
\begin{eqnarray}
&&M=\frac{3}{2}\pi \nu^2 + \frac{\pi}{8\nu^2} q^2 
 + O(q^4) \ , \nonumber \\
&& J=\pi \nu^3 \ (\mbox{for any order}) \ , \ \ \ Q=\pi q \ , 
\nonumber \\
&&\mu_{\rm {mag}}=\pi \nu q - \frac{\pi}{18 \nu^3} q^3 + O( q^5) \
. \label{mag_mom}
\end{eqnarray}
Thus the gyromagnetic ratio $g$ is obtained as
\begin{equation}
g=
3 + \frac{1}{12 \nu^4} q^2 +
  O(q^4) \ , \label{g_factor_extremal}
\end{equation}
which clearly differs from the lowest order value
$g^{(1)}=3$.
Expressing the $g$-factor in terms of the
charge to mass ratio $Q/M$, we obtain to this order
\begin{equation}
\frac{g}{3}=  1 + \frac{1}{16} \left(\frac{Q}{M}\right)^2 + \dots \ .
\label{g_5}
\end{equation}

With the event horizon located at $r=r_{H}$, where
\begin{equation}
r_{H} = \sqrt{2} \nu + \frac{\sqrt{2}}{24 \nu^3} q^2 
+ O( q^4) \ , \label{hor_rad}
\end{equation}
the horizon properties are given by
\begin{eqnarray}
&&\Omega = \frac{1}{2\nu} - \frac{1}{24 \nu^5} q^2 
+ O(q^4) \ , 
\nonumber \\
%
&&A_{H} = 8 \pi^2 \nu^3  + O(q^6) \ ,
\nonumber \\
&&\Phi_{H}=\frac{1}{4\nu^2} q + \frac{1}{72 \nu^6} q^3 +
O(q^5) \ , \label{AH_and_PhiH}
\end{eqnarray}
and the surface gravity $\kappa_{\rm sg}$
vanishes for these extremal solutions.

\section{Charged black holes in odd dimensions}

In higher dimensions, the expansions become increasingly intricate.
In the Einstein-Maxwell case and for odd $N$ 
Einstein-Maxwell-Chern-Simons solutions,
we can take advantage of the charge conjugation symmetry,
which reduces the expansions considerably,
since charge conjugation symmetry allows only for even
resp.~odd powers in the perturbative parameter
for the metric resp.~the gauge potential.
Still the presence of the Chern-Simons term
represents a substantial complication for
the perturbative expansion. 
Moreover, we can take advantage of the fact, that the
Chern-Simons term is of order $N+1$ in the charge,
and thus contributes only with order $N$ or higher
to the perturbative expansion.

To obtain the perturbative solutions in $D$ dimensions, we proceed
as in 5 dimensions. We fix the angular momentum
for any perturbative order, 
and impose the extremality condition for all orders. 
With this choice and by employing a regularity condition
at the horizon, we fix all integration constants.
The Smarr formula Eq.~(\ref{smarr}) is then employed
as a consistency check of the solutions.

\subsection{7 dimensions}

In 7 dimensions, the Chern-Simons term is even with respect
to charge conjugation. Thus as for the Einstein-Maxwell case
the expansion of the metric functions contains only even
powers in the perturbative parameter,
while the gauge functions contain only odd powers.
In the equations of motion,
the Chern-Simons term modifies the Maxwell equations
starting with power $q^3$
and the Einstein equations starting with power $q^4$.
Thus to 3rd order, only the gauge potential is modified
as compared to the pure Einstein-Maxwell case.
Since because of charge conjugation symmetry
no 4th order terms appear in the gauge functions,
the expansion for the gauge functions holds up to 4th order.

Introducing the parameter $\nu$ for the extremal 
Myers-Perry solutions by 
\begin{equation}
\hM=\frac{27}{8}\nu^4, \quad\quad
\hJ=\frac{27}{8}\nu^5, 
\label{nu7}
\end{equation}
the perturbations up to 3rd order then read
\begin{eqnarray}
&&g_{tt} = -1+\frac{27\nu^{4}}{4r^{4}}-q^{2}\left[\frac{8}{5r^{8}}-\frac{32}{135r^{4}\nu^{4}}\right]+O(q^{4})  \ ,
    \nonumber \\
&&W=1-\frac{27}{4}\frac{\nu^4}{r^4}+\frac{27}{4}\frac{\nu^6}{r^6}-q^2\left[\frac {32}{135r^{4}\nu^{4}}+\frac {16}{45\nu^{2}r^{6}}-\frac{8}{5r^{8}}+{\frac {4\nu^{2}}{5r
^{10}}}
\right]
+O(q^4) \ ,
    \nonumber \\
&&V=\frac{27 \nu^{6}}{4r^4}-q^2 \left[\frac{8\nu^{2}}{5 r^8}+\frac{16}{45\nu^2 r^4}\right]   
+O(q^4) \ ,
\nonumber \\
&&B=\frac{27 \nu^5}{4r^4} - \frac{8\nu}{5 r^8}  q^2 
+O(q^4) \ ,
\nonumber \\
&&a_t=\frac{q}{r^4}+\frac{16}{164025 r^8\nu^{13}}(2 r^2+ 3\nu^2)\left\{2r^4(2r^2-3\nu^2)(\nu
  -20 {\tilde
    \lambda})\left[80\ln\left(1-\frac{3\nu^{2}}{2r^2}\right)\right. \right. \ \  
\nonumber \\
&& \left.\left. +\ln\left(1+\frac{3\nu^{2}}{r^2}\right)\right]
  +9\nu^3 (52 r^4 -36 r^2\nu^2 -27 \nu^4) \right\} q^3 
-\frac{64{\tilde \lambda}}{3645 r^{10} \nu^{11}}(104r^8+84 r^6 \nu^2 \nonumber \\
&& -162r^4\nu^4-81r^2\nu^6-54\nu^8)q^3 + O(q^5)  \ , \ \ \nonumber  \\
&&a_\vphi=\frac{-\nu q}{r^4}-\frac{16}{492075}(\nu-20{\tilde
  \lambda})\left\{2r^4 (8r^6-27\nu^6) \left[80\ln\left(1-\frac{3\nu^{2}}{2r^2}\right)\right. \right. \ \  
\nonumber \\
&& \left.\left. +\ln\left(1+\frac{3\nu^{2}}{r^2}\right)\right]
+9\nu^2(208r^8+168r^6\nu^2-486r^2\nu^6 -243\nu^8)\right\} q^3  \nonumber  \\
&&-\frac{128(3\nu-160{\tilde \lambda})}{18225 r^4 \nu^8} q^3 + O(q^5) \ .
  \label{solution-7dim}
\end{eqnarray}
As in 5 dimensions, the expansion is not purely polynomial in $1/r$,
but log-terms appear in the expansion, starting in 3rd order
in the gauge potential and 4th order in the metric
\cite{NavarroLerida:2007ez}.

The global physical quantities are given to this order by
\begin{eqnarray}
&&M=\frac{135}{64}\pi^2 \nu^4 + \frac{2\pi^2}{27\nu^4} q^2 
 + O(q^4) \ , \nonumber \\
&& J=\frac{27}{32}\pi^2 \nu^5 \ (\mbox{for any order}) \ , \ \ \ Q=\pi^2 q \ , 
\nonumber \\
&&\mu_{\rm {mag}}=\pi^2 \nu q - \frac{128\pi^2(3\nu +40{\tilde \lambda})}{18225 \nu^8} q^3 + O( q^5) \
, \label{mag_mom-7dim}
\end{eqnarray}
yielding for the gyromagnetic ratio $g$ 
\begin{equation}
g=5 + \frac{256(\nu-20{\tilde \lambda})}{3645 \nu^9} q^2 +
  O(q^4) \ , \label{g_factor_extremal-7dim}
\end{equation}
which again differs from the lowest order value
$g^{(1)}=5$.
Expressing the $g$-factor in terms of the
charge to mass ratio $Q/M$, we obtain to this order
\begin{equation}
\frac{g}{5}=  1 + \frac{1}{16}\left(1-\frac{20{\tilde \lambda}}{\nu}\right) \left(\frac{Q}{M}\right)^2 + \dots \ .
\label{g_7}
\end{equation}
Interestingly, the correction term quadratic in the charge to mass ratio
$Q/M$ for pure EM black holes (i.e., when ${\tilde \lambda}=0$) is the same as in $D=5$ dimensions.

The event horizon is located at $r=r_{\rm H}$, where
\begin{equation}
r_{\rm H} = \frac{\sqrt{6} \nu}{2} + \frac{32\sqrt{6}}{3645 \nu^7} q^2 
+ O( q^4) \ , \label{hor_rad-7dim}
\end{equation}
with horizon properties 
\begin{eqnarray}
&&\Omega = \frac{2}{3\nu} - \frac{256}{10935 \nu^9} q^2 
+ O(q^4) \ , 
\nonumber \\
&&A_{\rm H} =\frac{27\sqrt{2}\pi^3\nu^5}{8} + O(q^4) \ ,
\nonumber \\
&&\Phi_{\rm H}=\frac{4}{27\nu^4} q+\frac{1024(3\nu + 100{\tilde \lambda})}{492075\nu^{13}} q^3 +
O(q^5) \ , \label{AH_and_PhiH-7dim}
\end{eqnarray}
and the surface gravity $\kappa_{\rm sg}$
vanishes for these extremal solutions.

\subsection{$D$ dimensions}

In $D$ dimensions, the Chern-Simons term is even with respect
to charge conjugation when $N$ is odd ($D=7,11,\dots$). 
When $N$ is even ($D=9,13,\dots$)
only the discrete symmetry Eq.~(\ref{cs_sym1}) holds.
However, in the equations of motion,
the Chern-Simons term only modifies the Maxwell equations
starting with power $q^N$
and the Einstein equations starting with power $q^{N+1}$.
Thus to 3rd order, the same formulae hold
as in the pure Einstein-Maxwell case, when $N>3$.

For the general case of $D$ dimensions
we now give the expressions for the Einstein-Maxwell case,
which hold for the metric in 3rd order and for the gauge potential
in 4th order. 
These expressions then hold to the same order
for general Chern-Simons coupling in $D > 9$ dimensions,
while in $9$ dimensions the expansion for the gauge potential
holds only up to 3rd order and not to 4th order.

To obtain the perturbative expansion 
we introduce the parameter $\nu$ 
for the extremal Myers-Perry solutions in $D$ dimensions by 
\begin{equation}
\hat{M}=\frac{(D-1)^{\frac{(D-1)}{2}}\nu^{D-3}}{4(D-3)^{\frac{(D-3)}{2}}}\ ,
\quad\quad
\hat{J}=\frac{(D-1)^{\frac{(D-1)}{2}}\nu^{D-2}}{4(D-3)^{\frac{(D-3)}{2}}}\ .
\label{nu_gen}
\end{equation}
To fix all the constants, we again need to make use of the
regularity condition of the horizon,
and employ the Smarr relation as a consistency check.
The perturbative expansion for the metric
and the gauge functions is exhibited in Appendix A.

Comparing the perturbative expansion to the asymptotic behavior 
of the solutions, Eqs.~(\ref{ggtt}-\ref{aphiphi}), we obtain
the global properties of the black holes
\begin{eqnarray}\label{mass}
M =\frac{A(S^{D-2})}{64\pi G_{D}}\left[\frac{2\nu^{2(D-3)}(D-2)(D-1)^{D-1}+16q^{2}(D-3)^{D-2}}{\nu^{D-3}(D-1)^{\frac{D-1}{2}}(D-3)^{\frac{D-3}{2}}}\right]+O(q^{4}) \ ,
\end{eqnarray}
\begin{eqnarray}\label{charge}
Q =\frac{A(S^{D-2})(D-3)}{4\pi G_{D}}q \ ,
\end{eqnarray} 
\begin{eqnarray}\label{angmom}
J =\frac{A(S^{D-2})\nu^{D-2}(D-1)^{\frac{D-1}{2}}}{16\pi G_{D}(D-3)^{\frac{D-3}{2}}} \ ,
\end{eqnarray}
\begin{eqnarray}\label{magmom}
\mu_{mag} = \frac{A(S^{D-2})(D-3)}{4\pi G_{D}}\left[q\nu-\frac{4q^3(D-3)^{D-2}}{\nu^{2D-7}(D-2)^{2}(D-1)^{D-2}}\right]+O(q^{5}) \ .
\end{eqnarray}
In particular, the gyromagnetic ratio g is given by
\begin{eqnarray}\label{g2}
g = D-2+4q^{2}\frac{(D-3)^{D-1}}{(D-2)(D-1)^{D-1}\nu^{2(D-3)}}+O(q^{4}) \ .
\end{eqnarray}

Thus the previous observation \cite{NavarroLerida:2007ez} 
in Einstein-Maxwell theory generalizes
to arbitrary (higher) odd dimensions:
higher order perturbation theory 
modifies the gyromagnetic ratio
from its lowest order value $g^{(1)}=D-2$.
Consequently, the gyromagnetic ratio is not constant as in 4 dimensions,
but for equal-magnitude angular momenta
the $g$-factor increases with increasing (small) $Q/M$
in any odd dimension $D \ge 5$.

Expressing the $g$-factor of Einstein-Maxwell black holes
again in terms of the
charge to mass ratio $Q/M$, we obtain to this order
\begin{equation}
\frac{g}{D-2}=  1 + \frac{1}{16} \left(\frac{Q}{M}\right)^2 + \dots \ .
\label{g_D}
\end{equation}
Thus we realize, that the first correction term to $g/(D-2)$, 
i.e., the first correction term 
to the gyromagnetic ratio normalized by its lowest order value,
is universal for extremal Einstein-Maxwell 
black holes with equal-magnitude angular momenta.
It does not depend on the dimension.

\begin{figure}[tbp]
\epsfxsize=7cm
\centerline{\includegraphics[angle=0,width=92mm]{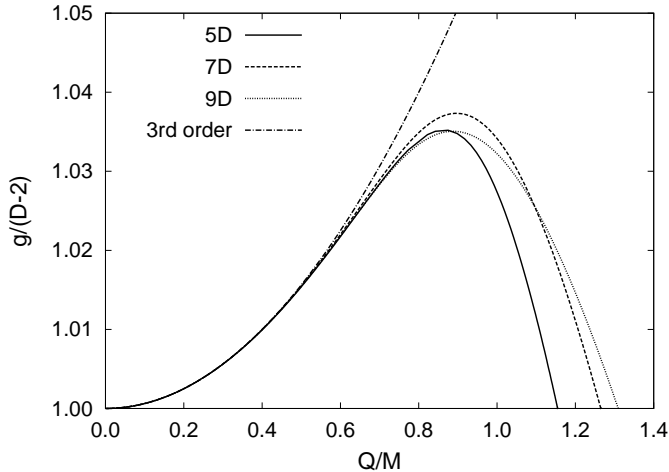}}
\caption{
The scaled gyromagnetic ratio $g/(D-2)$ versus
the charge to mass ratio $Q/M$
in $D = 5$ (solid line), $D = 7$ (dashed line),
and $D = 9$ (dotted line) dimensions:
exact numerical results versus the
3rd order perturbative (dot-dashed) results in Einstein-Maxwell theory.
}
\label{fig3}
\end{figure}

This is demonstrated explicitly in Fig.~\ref{fig3},
where the scaled gyromagnetic ratio $g/(D-2)$ is shown 
versus the charge to mass ratio $Q/M$ for $D=5$, 7 and 9 dimensions.
The range of validity of this 3rd order perturbative result
covers roughly half the domain of $Q/M$,
as comparison with numerical results demonstrates.
Close to the largest possible values of the charge to mass ratio,
i.e., close to the extremal Tangherlini solution,
however, perturbation theory in the angular momentum 
would be favourable, yielding the limiting value
$g/(D-2)=1$ \cite{Aliev:2005npa,Aliev:2006yk}. 

The event horizon of these black holes is located at
\begin{equation}
r_H=\sqrt{\frac{D-1}{D-3}}\nu+\frac{4(D-3)^{D-\frac{5}{2}}}{(D-2)(D-1)^{D-\frac{3}{2}}\nu^{2D-7}}q^2+O(q^{4}) \ . \label{r_H_odddim}
\end{equation}
The horizon angular velocities $\Omega$ and the horizon area $A_H$
are given by
\begin{eqnarray}\label{Omega}
&&\Omega = \frac{D-3}{\nu(D-1)}
-\frac{8q^2(D-3)^{D-1}}{\nu^{2D-5}(D-2)(D-1)^{D}}+O(q^{4}) \ ,
\nonumber \\
%
&&A_H=\frac{\sqrt{2}A(S^{D-2})(D-1)^{\frac{(D-1)}{2}}\nu^{D-2}}{2(D-3)^{\frac{(D-2)}{2}}}+O(q^{4}) \ ,
\end{eqnarray}
and the surface gravity $\kappa_{sg}$ vanishes
for these extremal solutions.



\section{Summary and Conclusion}

Focussing on odd dimensions,
where the Einstein-Maxwell action
can be supplemented with a Chern-Simons term,
we have obtained a set of extremal charged rotating black holes 
with equal-magnitude angular momenta,
that are asymptotically flat,
and whose horizon has spherical topology.

Our strategy for obtaining these solutions
was based on the perturbative method,
where we solved the equations of motion
up to the 3rd order in the perturbative parameter,
which we chose proportional to the charge.
In particular, we started from the
rotating black hole solutions in higher dimensions
\cite{Myers:1986un}.
We then considered the effect of adding a small amount
of charge 
to the solutions,
and evaluated how the perturbative parameter 
modified the physical properties of the solutions.

In 5 dimensions, we have derived in this way
the metric, the gauge potential,
and the physical properties of charged extremal rotating black holes
for arbitrary values of the Chern-Simons coupling constant $\tilde \lambda$.
For the supergravity value of $\tilde \lambda$ our 3rd order results
were found to agree with the analytically known solution up to this order,
showing the consistency of the perturbative approach.

Likewise, in higher odd dimensions, we have obtained the general 
3rd order perturbative solutions for 
arbitrary values of the Chern-Simons coupling constant 
and investigated their physical properties.
In the case of charge conjugation symmetry
the expansion in the charge simplifies,
since only even resp.~odd terms can enter the expansion of
the metric resp.~gauge potential.
Consequently, the expressions for the gauge potential hold
to 4th order.

For the $g$-factor of
the extremal Einstein-Maxwell-Chern-Simons black holes 
we obtained the following expansions 
in terms of the charge to mass ratio $Q/M$
\begin{equation}
\displaystyle D=5: \ \ \ \ \frac{g}{3} = 1 - \frac{3{\tilde \lambda}}{2} \frac{Q}{M} +
\frac{1}{16} (1+24 {\tilde \lambda}^2) \left(\frac{Q}{M}\right)^2 +
O\left(\left(\frac{Q}{M}\right)^3 \right) \ , \label{D_5}
\end{equation}
\begin{equation}
\displaystyle D=7: \ \ \ \ \frac{g}{5} = 1 + \frac{1}{16}
\left(1-20\frac{{\tilde \lambda}}{\nu} \right)  \left(\frac{Q}{M}\right)^2 +
O\left(\left(\frac{Q}{M}\right)^3 \right) \ , \label{D_7}
\end{equation}
\begin{equation}
\displaystyle D>7: \ \ \ \ \frac{g}{D-2} = 1 + \frac{1}{16}\left(\frac{Q}{M}\right)^2 +
O\left(\left(\frac{Q}{M}\right)^3 \right) \ . \label{D_large}
\end{equation}

Restricting to Einstein-Maxwell black holes then yields the result
that in 3rd order perturbation theory
the $g$-factor 
deviates from the 1st order result $g^{(1)}=D-2$
\cite{Aliev:2005npa,Aliev:2006yk}.
Indeed, the 3rd order result shows
that the $g$-factor of extremal black holes
with equal-magnitude angular momenta increases with increasing charge
in any odd dimension.
Moreover, when expressing the $g$-factor in terms of the
charge to mass ratio $Q/M$, we obtain to this order
\begin{equation}
\frac{g}{D-2}=  1 + \frac{1}{16} \left(\frac{Q}{M}\right)^2 + \dots \ .
\label{g_DC}
\end{equation}
Thus the correction term to $g/(D-2)$
is universal for 
these black holes.
It does not depend on the dimension.

Comparison with available analytical and numerical results demonstrated that
the range of validity of these 3rd order perturbative results
covers a considerable portion 
of the allowed range of $Q/M$ values,
whose boundaries are determined by the extremal Tangherlini solution.
In the vicinity of the extremal Tangherlini solution,
however, perturbation theory in the angular momenta
should be favourable. 
An extension to higher order
of this perturbative approach in the angular momenta
\cite{Aliev:2005npa,Aliev:2006yk}
might well supplement the present higher order perturbative approach,
and yield a good approximation
for the larger values of the charge to mass ratio.

Here we have only considered extremal charged rotating
black holes in higher dimensions. 
The generalization of the present work to the non-extremal case 
is currently under investigation 
and will be addressed elsewhere.
However, some results in the pure Einstein-Maxwell case
5 dimensions have already been given in
\cite{NavarroLerida:2007ez}.

The Myers-Perry black holes can be generalized
to allow for a cosmological constant \cite{Gibbons:2004js}.
But rotating AdS Einstein-Maxwell black holes are not known it closed form.
Charged rotating AdS Einstein-Maxwell
black holes have been considered numerically 
\cite{Kunz:2007jq,Brihaye:2008br} as well as in
lowest order perturbation theory \cite{Aliev:2006tt,Aliev:2007qi}
also in the presence of a Gau\ss -Bonnet term
\cite{Kim:2007iw}.
Generalization of these calculations to higher order
perturbation theory will be an important next step.

\appendix

\section{} 

Here we give the perturbative expressions 
for the metric and the gauge potential in
Einstein-Maxwell theory for general odd $D$.
The solutions up to 3rd order read
\begin{eqnarray}\label{ggtt}
g_{tt} &=& -1+\frac{(D-1)^{\frac{(D-1)}{2}}\nu^{D-3}}{2(D-3)^{\frac{(D-3)}{2}}r^{D-3}}\nonumber \\
&&-q^{2}\left[\frac{2(D-3)}{(D-2)r^{2(D-3)}}-\frac{4(D-3)^{\frac{D-1}{2}}}{(D-2)(r\nu)^{D-3}(D-1)^{\frac{D-1}{2}}}\right]+O(q^{4}) \  ,
\end{eqnarray}
\begin{eqnarray}\label{WW}
W &=&1-\frac{(D-1)^{\frac{(D-1)}{2}}\nu^{D-3}}{2(D-3)^{\frac{(D-3)}{2}}r^{D-3}}+\frac{(D-1)^{\frac{(D-1)}{2}}\nu^{D-1}}{2(D-3)^{\frac{(D-3)}{2}}r^{D-1}} -\frac{2q^{2}}{D-2}\Bigg{\{}\frac{2(D-3)^{\frac{(D-1)}{2}}}{(r\nu)^{D-3}(D-1)^{\frac{(D-1)}{2}}}\nonumber \\
&&+\frac{(D-3)^{\frac{(D-3)}{2}}}{\nu^{D-5}r^{D-1}(D-1)^{\frac{(D-3)}{2}}}+\frac{(D-5)\nu^{2}}{r^{2(D-2)}}-\frac{D-3}{r^{2(D-3)}}\Bigg{\}}+O(q^{4}) \ ,
\end{eqnarray}
\begin{eqnarray}\label{VV}
V &=& \frac{(D-1)^{\frac{(D-1)}{2}}\nu^{D-1}}{2(D-3)^{\frac{(D-3)}{2}}r^{D-3}}\nonumber \\
&&-q^{2}\left[\frac{2(D-3)\nu^2}{(D-2)r^{2(D-3)}}+\frac{4(D-3)^{\frac{D-3}{2}}}{(D-2)(D-1)^{\frac{D-3}{2}}\nu^{D-5}r^{D-3}}\right]+O(q^{4}) \ ,
\end{eqnarray}
\begin{eqnarray}\label{BB}
B = \frac{(D-1)^{\frac{(D-1)}{2}}\nu^{D-2}}{2(D-3)^{\frac{(D-3)}{2}}r^{D-3}}-\frac{2\nu(D-3)q^{2}}{(D-2)r^{2(D-3)}}+O(q^{4}) \ ,
\end{eqnarray}
\begin{eqnarray}\label{a00}
a_{0} &=& \frac{q}{r^{D-3}}+q^3\Bigg{\{}\int\frac{1}{r^{3(D-2)}}\Bigg{\{}\frac{\nu^{2}(D-1)(D-3)^{D-1}r^{2(D-2)}}{(D-2)2^{(D-6)}} \times I_{1}\times I_{2}+\frac{8I_{2}\times S_{3}}{(D-3)^{3}}\nonumber \\
&&-\frac{16}{3}\frac{\ln(\frac{D-3}{2})(D-3)^{D-1}r^{2(D-2)}}{\nu^{2(D-3)}(D-2)(D-1)^{D-2}}\Bigg{\}}dr-\frac{16}{3}\frac{\ln(\frac{D-3}{2})(D-3)^{\frac{3D-7}{2}}}{\nu^{3(D-3)}(D-2)(D-1)^{\frac{3D-7}{2}}}\Bigg{\}}\nonumber \\
&&+O(q^{5}) \ ,
\end{eqnarray}
\begin{eqnarray}\label{aphiphi}
a_{\varphi}&=&\frac{-\nu q}{r^{D-3}}+q^{3}r^{2}\Bigg{\{}\int\frac{1}{\left[\nu^{2D-7}r^{4D-5}-\frac{\nu^{3D-10}(D-1)^{\frac{D-1}{2}}r^{3D-2}}{2(D-3)^{\frac{D-3}{2}}}+\frac{(D-1)^{\frac{D-1}{2}}\nu^{3D-8}r^{3D-4}}{2(D-3)^{\frac{D-3}{2}}}\right]}\nonumber \\
&&\Bigg{\{}\frac{(D-1)^{2}(D-3)^{D-2}(\nu r)^{2(D-2)}}{2^{D-6}(D-2)}\left(r^{D-1}-\frac{\nu^{D-3}(D-1)^{\frac{D-1}{2}}r^{2}}{2(D-3)^{\frac{D-3}{2}}}+\frac{\nu^{D-1}(D-1)^{\frac{D-1}{2}}}{2(D-3)^{\frac{D-3}{2}}}\right) \nonumber \\
&&\times I_{1} \times I_{2}-S_{1}\times I_{2}+S_{2}\Bigg{\}}dr+\frac{16}{3}\frac{\ln(\frac{D-3}{2})(D-3)^{\frac{3D-5}{2}}}{\nu^{3D-8}(D-2)(D-1)^{\frac{3D-5}{2}}}\Bigg{\}}+O(q^{5}) \ ,
\end{eqnarray}
where in the above equations $I_{1}$ , $I_{2}$, $S_{1}$, $S_{2}$ and $S_{3}$ are

\begin{eqnarray}\label{I1}
I_{1} = \int\frac{1}{r^{D-2}}-\frac{\nu^{D-3}(D-1)^{\frac{D-1}{2}}}{2(D-3)^{\frac{D-3}{2}}r^{2D-5}}+\frac{(D-1)^{\frac{D-1}{2}}\nu^{D-1}}{2(D-3)^{\frac{D-3}{2}}r^{2D-3}}dr \ ,
\end{eqnarray}

\begin{eqnarray}\label{I2}
I_{2} = \int\frac{r^{D-2}}{\left(\frac{(D-3)r^2}{2}-\frac{(D-1)\nu^{2}}{2}\right)^{4}\left(\sum^{\frac{D-5}{2}}_{i=0}(i+1)\nu^{2i}(\frac{D-1}{2})^{i}(\frac{D-3}{2})^{\frac{D-2i-7}{2}}r^{D-2i-5}\right)^{2}}dr \ ,
\end{eqnarray}

\begin{eqnarray}\label{S1}
S_{1} &=& \frac{\nu^{2(D-3)}2^{\frac{D+3}{2}}(D-1)S_{3}}{(D-3)^{\frac{D+5}{2}}}\left(\frac{(D-3)r^2}{2}-\frac{(D-1)\nu^{2}}{2}\right)^{2}\nonumber \\
&&\left(\sum^{\frac{D-5}{2}}_{i=0}(i+1)\nu^{2i}(\frac{D-1}{2})^{i}(\frac{D-3}{2})^{\frac{D-2i-7}{2}}r^{D-2i-5}\right) \ ,
\end{eqnarray}

\begin{eqnarray}\label{S2}
S_{2} &=&\frac{4(D-3)^{D-3}r^{3D-5}}{(D-1)^{D-4}(D-2)^{2}}-\frac{8}{3}\frac{\nu^{D-3}\ln(\frac{D-3}{2})(D-3)^{\frac{D-1}{2}}r^{2(D-1)}}{(D-2)(D-1)^{\frac{D-5}{2}}}+\frac{4\nu^{D-3}(D-3)^{\frac{D-1}{2}}r^{2(D-1)}}{(D-2)(D-1)^{\frac{D-3}{2}}}\nonumber \\
&&+\frac{4\nu^{2(D-2)}(D-1)r^{D-1}}{(D-2)^{2}}-\frac{8(\frac{D-3}{D-1})^{D-3}r^{3D-5}}{(D-2)}-\frac{4\nu^{D-1}(D-3)^{\frac{D-3}{2}}r^{2(D-2)}}{(D-2)(D-1)^{\frac{D-5}{2}}}\nonumber \\
&&+\frac{8}{3}\frac{\nu^{D-1}\ln(\frac{D-3}{2})(D-3)^{\frac{D-1}{2}}r^{2(D-2)}}{(D-2)(D-1)^{\frac{D-5}{2}}}+\frac{16}{3}\frac{\ln(\frac{D-3}{2})(D-3)^{D-2}r^{3D-5}}{(D-2)(D-1)^{D-3}} \ ,
\end{eqnarray}

\begin{eqnarray}\label{S3}
S_{3}&=&\frac{(D-3)^{\frac{3D-1}{2}}r^{2(D-2)}}{2^{D-1}\nu^{D-5}(D-2)^{2}(D-1)^{\frac{D-7}{2}}}-\frac{\nu^{2}(D-1)(D-3)^{D+1}r^{D-1}}{2^{D-3}(D-2)}\nonumber \\
&&+\frac{\nu^{D-1}(D-1)^{\frac{D+1}{2}}(D-3)^{\frac{D+5}{2}}r^{2}}{2^{D-1}(D-2)}-\frac{\nu^{D+1}(D-1)^{\frac{D+1}{2}}(D-3)^{\frac{D+7}{2}}}{2^{D-1}(D-2)^{2}}
\ .
\end{eqnarray}

\acknowledgments{FNL gratefully acknowledges Minis\-terio de Ciencia e
  Innovaci\'on of Spain for financial support under project FIS2009-10614.}


\end{document}